\PassOptionsToPackage{x11names,rgb,html,table}{xcolor}
\documentclass[sigplan,nonacm,authorversion,screen,british,final]{acmart}
\pdfoutput=1

\usepackage{etoolbox}

\usepackage[T1]{fontenc}
\usepackage[utf8]{inputenc}
\usepackage{microtype}

\usepackage[x11names,rgb,html]{xcolor}

\usepackage[british]{babel}

\usepackage{amsmath}

\usepackage{float}
\usepackage[section]{placeins}

\usepackage[inline]{enumitem}

\usepackage{hyperref}
\usepackage{nameref}

\usepackage[capitalise,nameinlink,noabbrev]{cleveref}

\AtEndPreamble{\hypersetup{
	hidelinks,
	final,
}}

\usepackage[final]{listings}

\lstdefinelanguage{jolie}{
  morekeywords={as,csets,type,configType,raw,any,undefined,void,default,
  if,for,while,spawn,foreach,else,define,main,include,from,
  constants,inputPort,outputPort,interface,execution,cset,
  nullProcess,RequestResponse,OneWay,throw,throws,install,
  scope,embedded,init,synchronized,global,is_defined,
  is_int,is_bool,is_long,is_string,bool,long,int,string,
  double,undef,with,location,protocol,interfaces,aggregates,courier,
  redirects,import,embed,service,binding,foreign,InputPort,OutputPort,implementation},
  morekeywords=[2]{>>,select,match},
  sensitive,%
	morecomment=[l]//,%
	morecomment=[s]{/*}{*/},%
	morestring=[b]",%
  morestring=[b]',%
  classoffset=1, %
  otherkeywords={;,|,@},
  classoffset=0,%
  }[keywords,comments,strings]%

\lstdefinelanguage{docker}{
  morekeywords={FROM,COPY,CMD},
  sensitive,%
	morecomment=[l]//,%
	morecomment=[s]{/*}{*/},%
	morestring=[b]",%
  morestring=[b]',%
  classoffset=1, %
  otherkeywords={;,|,@},
  classoffset=0,%
  }[keywords,comments,strings]%
  
\lstdefinelanguage{compose}{
  morekeywords={version,services,build},
  sensitive,%
	morecomment=[l]//,%
	morecomment=[s]{/*}{*/},%
	morestring=[b]",%
  morestring=[b]',%
  classoffset=1, %
  otherkeywords={;,|,@},
  classoffset=0,%
  }[keywords,comments,strings]%

\definecolor{sbase03}{HTML}{002B36}
\definecolor{sbase02}{HTML}{073642}
\definecolor{sbase01}{HTML}{586E75}
\definecolor{sbase00}{HTML}{657B83}
\definecolor{sbase0}{HTML}{839496}
\definecolor{sbase1}{HTML}{93A1A1}
\definecolor{sbase2}{HTML}{EEE8D5}
\definecolor{sbase3}{HTML}{FDF6E3}
\definecolor{syellow}{HTML}{B58900}
\definecolor{sorange}{HTML}{CB4B16}
\definecolor{sred}{HTML}{DC322F}
\definecolor{smagenta}{HTML}{D33682}
\definecolor{sviolet}{HTML}{6C71C4}
\definecolor{sblue}{HTML}{268BD2}
\definecolor{scyan}{HTML}{2AA198}
\definecolor{sgreen}{HTML}{859900}

\usepackage[scaled=.75]{FiraMono}
\lstdefinestyle{solarized-light}{
  frame=none,
  breaklines=true,
  showstringspaces=false,
  tabsize=2,
  columns=fixed,
  mathescape=true,
  extendedchars=true,
  backgroundcolor=\color{sbase3},
  keywordstyle=\bfseries\color{sbase01},
  keywordstyle = [2]{\bfseries\color{sblue}},
  stringstyle=\color{sviolet},
  numberstyle=\color{sviolet},
  identifierstyle=\color{sbase03},
  commentstyle=\color{sgreen},
  basicstyle=\color{sbase03}\ttfamily
}

\usepackage[most]{tcolorbox}
\tcbuselibrary{listings}

\NewDocumentCommand{\newlang}{o m m m}{%
  \IfNoValueTF{#1}{\def\n{#2}}{\def\n{#1}}
  \expandafter\NewDocumentCommand\csname \n listing\endcsname{s O{} O{}}{%
	  \def\WithoutTitle{\tcblisting{
	          enhanced, %
	          before skip=\abovedisplayskip,
	          after  skip=\belowdisplayskip,
	          sharp corners=all,
	          boxrule=2pt,
	          boxsep=-.7em,
	          colframe=#4,
	          colback=sbase3,
	          title={},
	          listing only,
	          listing options={language=#2,numbers=none,style=solarized-light,##3},
	          ##2
 	  	}}%
	  	\def\withTitle{\tcblisting{
	          enhanced, %
	          before skip=\abovedisplayskip,
	          after  skip=\belowdisplayskip,
	          sharp corners=all,
	          boxrule=2pt,
	          boxsep=-.7em,
	          colframe=#4,
	          colback=sbase3,
	          detach title,
	          finish={\node[anchor=north east, font=\footnotesize\itshape,
	          text=sbase3,fill=#4] at (frame.north east) {#3};},
	          title={},
	          listing only,
	          listing options={language=#2,numbers=none,style=solarized-light,##3},
	          ##2
	    }}%
	  \IfBooleanTF{##1}{\WithoutTitle}{\withTitle}}
  \expandafter\def\csname end\n listing\endcsname{\endtcblisting\noindent}
  \expandafter\def\csname\n\endcsname{\lstinline[language=#2,style=solarized-light]}
}

\newlang{jolie}{Jolie}{syellow}
\newlang[dockerfile]{docker}{Dockerfile}{scyan}
\newlang{compose}{Docker Compose}{sviolet}

\title{Sliceable Monolith: Monolith First, Microservices Later}

\hypersetup{
	pdftitle={Sliceable Monolith: Monolith First, Microservices Later},
  pdfauthor={Fabrizio Montesi, Marco Peressotti, Valentino Picotti}
}

\author{Fabrizio Montesi}
\orcid{0000-0003-4666-901X}
\affiliation{
  \department{Department of Mathematics and Computer Science}
  \institution{University of Southern Denmark}
  \streetaddress{Campusvej 55}
  \city{Odense}
  \postcode{5230}
  \country{Denmark}
}
\email{fmontesi@imada.sdu.dk}

\author{Marco Peressotti}
\orcid{0000-0002-0243-0480}
\affiliation{
  \department{Department of Mathematics and Computer Science}
  \institution{University of Southern Denmark}
  \streetaddress{Campusvej 55}
  \city{Odense}
  \postcode{5230}
  \country{Denmark}
}
\email{peressotti@imada.sdu.dk}

\author{Valentino Picotti}
\affiliation{
  \department{Department of Mathematics and Computer Science}
  \institution{University of Southern Denmark}
  \streetaddress{Campusvej 55}
  \city{Odense}
  \postcode{5230}
  \country{Denmark}
}
\email{picotti@imada.sdu.dk}

\begin{document}

\begin{abstract}
  We propose Sliceable Monolith, a new methodology for developing microservice
  architectures and perform their integration testing by leveraging most of the
  simplicity of a monolith: a single codebase and a local execution environment
  that simulates distribution. Then, a tool compiles a codebase for each
  microservice and a cloud deployment configuration. The key enabler of our
  approach is the technology-agnostic service definition language offered by
  Jolie.
\end{abstract}

\maketitle

\section{Introduction}
\label{sec:introduction}
Microservices represent the prominent software paradigm for building distributed
applications that strive for scalability, maintainability, and tight
development and deployment cycles~\cite{Detal17}. Microservices enforce strong
boundaries and interact by message passing, leading to modular and independently
executable software components.
However, they require dealing with multiple codebases (one per
microservice), making prototyping and testing more challenging compared to
a monolith---a standard application that consists of a single executable.

The complexity introduced by microservices can easily outweigh their benefits
and, especially when it comes to greenfield project development, experts have
mixed opinions on whether to start with microservices or with a
monolith~\cite{FowlerMonolithFirst,TilkovDontMonolith,NewmanGreenfield}.
Thus we ask: Can we recover some of the simplicity of monoliths in the
development of microservices? A positive answer would contribute to making the
greenfield development of microservice systems more approachable, which is
important because migrating monoliths to
microservices is difficult~\cite{TilkovDontMonolith}.

\begin{figure*}
\begin{center}
\includegraphics[width=\textwidth]{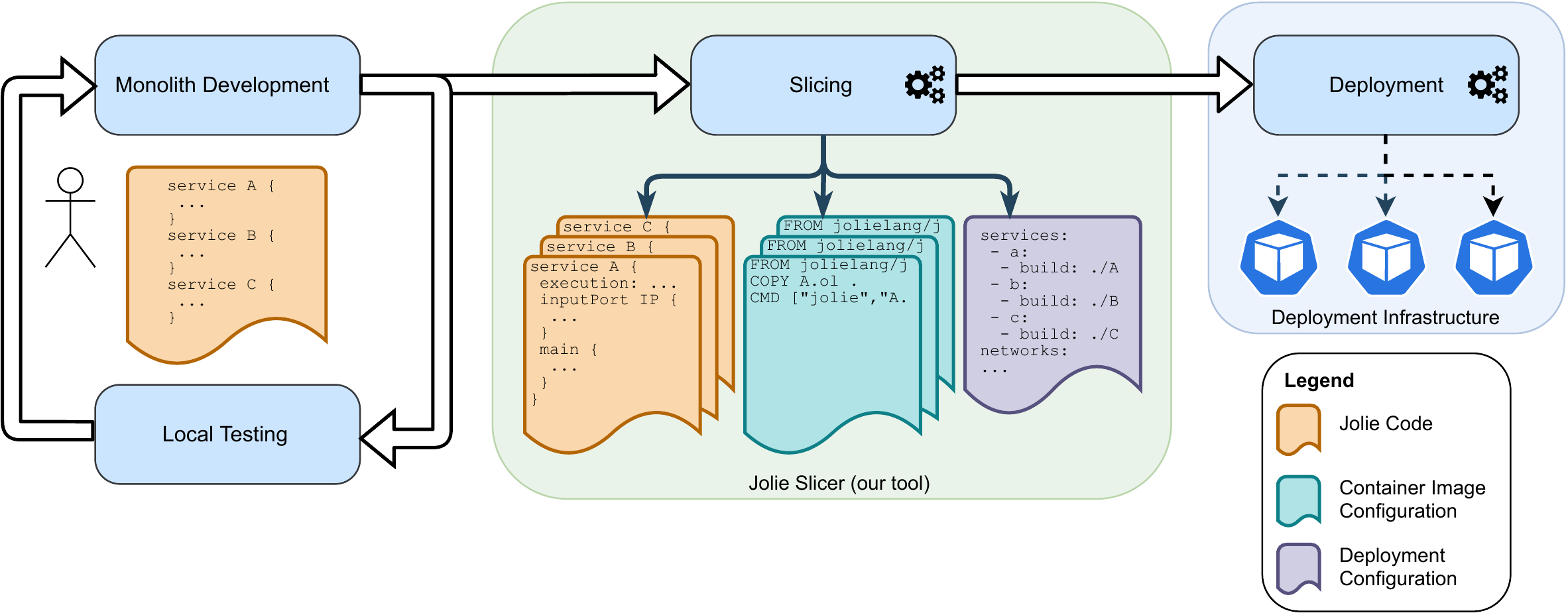}
\end{center}
\caption{The development methodology of Sliceable Monolith.}
\label{fig:methodology}
\end{figure*}

In this article, we propose a new development methodology whereby an entire
microservice architecture has a single codebase. Thus, our approach reduces
drastically the complexity of reaching a working prototype to iterate on. We
depict our methodology in \cref{fig:methodology} and outline it in the
following.

The main artifact in the codebase is a ``sliceable monolith'': the definition of a microservice system that looks like a monolith, but where all components are
enforced to be services with clear boundaries and data models (e.g., the
structures of Data Transfer Objects). We achieve these features by using the
Jolie programming language~\cite{MGZ14}. Jolie enforces linguistically some best practices for microservice development, e.g., interaction among components happens
necessarily through formally-defined service interfaces. Thanks to the built-in
facilities of the Jolie interpreter, the application can then be tested locally
straight away, enabling fast refinement cycles of the prototype.

The structure of a sliceable monolith make it possible to automatically extract
the implementation of each microservice into its own codebase. We implement this
procedure with an automatic \emph{slicer} tool (called Jolie Slicer),
emphasising the fact that the sliceable monolith is cut alongside the sharp
boundaries of the microservices. Our slicer tool
also produces the necessary configuration for the containerisation and
distributed deployment of the microservice system on the cloud.
At this point, developers are free to choose between iterating on the sliceable
monolith codebase, or to start developing some (even all) of the microservices
independently. The technology-agnostic nature of Jolie interfaces makes it
possible to mix different languages for the implementation of each microservice
(Jolie currently supports its own behavioural language, Java, and JavaScript,
with a plug-in architecture for adding more~\cite{MGZ14}).

\section{A Use Case from Smart Cities}

In this section, we present our development methodology with a running example based on (the relevant parts of) the implementation of a use case from smart cities.
We model a scenario in which a microservice architecture manages a set of private
parking areas whose owners, in agreement with the power grid operators, have
decided to share their charging stations in exchange for monetary incentives.
The end-users of the application are owners of electrical vehicles looking for
available charging stations near a given location. The described scenario is
borrowed from \cite{Rademacher2021,puls},
where microservices interact according to the Domain Event pattern. The application follows the CQRS pattern (Command Query Responsibility Segregation): two services, one for querying data (\texttt{QuerySide}) and the other for updating data (\texttt{CommandSide}) interact indirectly through an event store service (\texttt{EventStore}).

\subsection{Development and Local Testing}
Service \texttt{QuerySide} allows clients to obtain information about parking spaces equipped with charging stations, performing queries based on geolocation. Service \texttt{CommandSide} offers an API for updating information about parking spaces. Service \texttt{EventStore} supports the coordination of the other two services by offering an API for event-driven communication.

Following the Sliceable Monolith approach, in this subsection we define the interfaces (data models and APIs) and implementations of these services in a single codebase.

We start by writing a Jolie program (\texttt{smart-city.ol}) that defines a service block for each one of our services:
\begin{jolielisting}[breakable=true]
service QuerySide( config ) { ... }
service CommandSide( config ) { ... }
service EventStore( config ) { ... }
\end{jolielisting}
Each service is parameterised on an externally-provided configuration, called \texttt{config}.

We show the most interesting parts of the definition of service \texttt{CommandSide}, which is the most involved.
First, we use Jolie data types to define the data model of the messages that the service exchanges. These include parking area identifiers and structures for providing information about each area: name, the time period in which it is available, the speed supported by the charging station in the area, and geolocation.
\begin{jolielisting}[breakable=true]
type PAID:long // Parking Area IDentifier
type ParkingArea {
	id:PAID
	info:ParkingAreaInformation
}
type ParkingAreaInformation {
	name:string
	availability*:TimePeriod
	chargingSpeed:ChargingSpeed
	geolocation:Location
}
\end{jolielisting}

Using the data model, we build the API of \texttt{CommandSide} as a Jolie interface that comprises three RPCs for creating, updating, and deleting parking areas (\texttt{RequestResponse} is Jolie for RPC):
\begin{jolielisting}[breakable=true]
interface CommandSideInterface {
RequestResponse:
 createParkingArea( ParkingAreaInformation )( PAID ),
 updateParkingArea( ParkingArea )( string ),
 deleteParkingArea( PAID )( string )
}
\end{jolielisting}

In the definition of service \texttt{CommandSide}, we offer the API that we have just defined to clients (through a Jolie \texttt{inputPort}) and we declare a dependency towards the \texttt{EventStore} service (through a Jolie \texttt{outputPort}). The locations at which these communication ports should be deployed at are parameters that we get from the externally-provided configuration.
\begin{jolielisting}[breakable=true]
/* ... data types and API definitions ... */
service CommandSide( config:Configuration ) {
	execution: concurrent
	inputPort InputCommands {
		location: config.CommandSide.location
		protocol: http { format = "json" } 
		interfaces: CommandSideInterface
	}
	outputPort EventStore {
		location: config.EventStore.location
		protocol: http { format = "json" } 
		interfaces: EventStoreInterface
	}
	main { /* business logic implementation */ }
}
\end{jolielisting}
Note that data types and APIs appear outside of the \jolie{service} block that implements \texttt{CommandSide}. This allows us to share these same types across the implementations of all services, which aids in keeping the data models of interacting services consistent. For example, we can conveniently reuse the type \texttt{ParkingArea} in the data types of events that can be exchanged through the \texttt{EventStore} service. This convenience comes at zero cost: our slicer tool (next subsection) uses a dependency analysis to produce optimised code for each microservice.

Externally-provided configurations for Jolie services can be given as JSON files.
By using a JSON file that provides locations at the local host, we can test the entire architecture locally, both with unit and integration tests. While the latter is typically problematic in general, in our case it is simply a matter of writing a few lines of Jolie code that can be run locally. For example, the following test checks that deleting a parking area triggers the right event notification from \texttt{EventStore}.
\begin{jolielisting}[breakable=true]
subscribe@EventStore( {
	location = testLocation
	topics[0] = "PA_DELETED"
} )( res )
deleteParkingArea@CommandSide( 123L )()
notify( event )
if( event.type != "PA_DELETED" || event.id != 123L )
	throw( AssertionFailed )
\end{jolielisting}

\subsection{Slicing and Deployment}

To switch to a distributed architecture, we write another configuration file
(\texttt{deploy.json}), where locations are abstract DNS names (resolved by Docker, in our concrete deployment). We then run our Jolie Slicer tool with the following command.
\definecolor{slicing}{RGB}{185,213,171} %
\begin{tcblisting}{
	enhanced, 
	before skip=\abovedisplayskip,
	after  skip=\belowdisplayskip,
	sharp corners=all,
	boxrule=0pt,
	boxsep=-.7em,
	colframe=slicing,
	colback=sbase3,
	title={},
	listing only,
	listing options={language=bash,numbers=none,style=solarized-light}}
jolie-slicer --config deploy.json smart-city.ol
\end{tcblisting}
This produces an output directory that contains a
Docker Compose file for cloud deployment and a subfolder for each service. Each subfolder contains
the code of its respective service alongside a \texttt{Dockerfile} that
instructs Docker on how the service should be containerised, like the following.

\begin{dockerfilelisting}[breakable=true]
FROM jolielang/jolie
COPY CommandSide.ol .
COPY ../deploy.json .
CMD ["jolie", "-p", "deploy.json", "CommandSide.ol"]
\end{dockerfilelisting}

The generated \texttt{docker-compose.yml} file is compatible with Docker Swarm.
It can be used as is, but the programmer can also refine it, e.g., with the
desired load balancing configuration:

\begin{composelisting}[breakable=true]
services:
  commandside:
		build: ./commandside
    deploy:
      replicas: 1
  ...
\end{composelisting}

From a Docker Swarm manager node, the following command will deploy the entire
architecture as a composition of independent microservices, as expected.
\definecolor{deployment}{RGB}{205,217,234} %
\begin{tcblisting}{
	enhanced, 
	before skip=\abovedisplayskip,
	after  skip=\belowdisplayskip,
	sharp corners=all,
	boxrule=0pt,
	boxsep=-.7em,
	colframe=deployment,
	colback=sbase3,
	title={},
	listing only,
	listing options={language=bash,numbers=none,style=solarized-light}}
docker stack deploy -c docker-compose.yml smartcity
\end{tcblisting}

\section{Conclusion}
\label{sec:conclusions}

When developing a new service system, developers have to choose whether to start
with a monolith or jump directly to microservices. The first choice leads to
quick prototyping, but poses the risk of discarding the monolith entirely when
migrating to
microservices~\cite{FowlerMonolithFirst}. The
second choice gives immediately a more flexible architecture, but at the cost of
severely slowing down early development~\cite{Newman2015,TilkovDontMonolith}.

Sliceable Monolith is a new middle ground between these approaches, which retains
some of the simplicity of developing a monolith and automates the migration to a
full-fledged distributed system of microservices.

We have omitted the business logic
implementations of our services in the example; they are not surprising, and apply typical best practices (especially statelessness, where good scalability is desired).
Our methodology abstracts from the technologies used for such implementations. Our tool already supports different technologies through Jolie, which has a plug-in architecture for implementing the business logics of services (the ``\jolie{main}'' block) in other languages, like Java and JavaScript~\cite{MGZ14}.
Our tool currently supports Docker Swarm for cloud deployment, but in principle our methodology applies also to other technologies, like Kubernetes, Pulumi, or Terraform.

Our approach gives a first positive answer to the question in the introduction.
It joins multiparty languages, an emerging class of distributed programming paradigms~\cite{GMPRSW21}.
Interesting future work includes the systematic study of its impact.

\begin{acks}
  We thank Florian Rademacher for useful discussions about the use case.
  Work partially supported by Independent Research Fund Denmark, grant no.~0135-00219.
\end{acks}

\bibliography{biblio}

\end{document}